\documentclass[prd,twocolumn,superscriptaddress,floatfix]{revtex4}     

\usepackage{graphicx,psfrag}
\usepackage{mathrsfs}
\usepackage{amsmath,amsfonts,amssymb}

\usepackage{epsfig}
\usepackage{xcolor}
\usepackage{float}
\usepackage{ulem}

\usepackage{ifthen}

\newcommand{\be}{\begin{equation}}
\newcommand{\ee}{\end{equation}}
\newcommand{\bea}{\begin{eqnarray}}
\newcommand{\eea}{\end{eqnarray}}
\newcommand{\bel}{\begin{align}}
\newcommand{\eel}{\end{align}}

\newcommand\mnras{\ref@jnl{MNRAS}}

\def\bam{\textsc{BAM}}

\def\Msun{M_{\odot}}

\begin{document}

\title{Eccentric binary neutron star mergers}

\newcommand{\Jena}
{Theoretisch-Physikalisches Institut, Friedrich Schiller Universit\"at Jena,
Max-Wien-Platz 1, 07743~Jena, Germany}
\newcommand{\Princeton}
{Department of Physics, Princeton University, Princeton, NJ 08544, USA}

\author{Roman \surname{Gold}}
\affiliation{\Jena}
\affiliation{\Princeton}
\affiliation{Department of Physics, University of Illinois at Urbana-Champaign, Urbana, Illinois 61801}

\author{Sebastiano \surname{Bernuzzi}}
\affiliation{\Jena}

\author{Marcus \surname{Thierfelder}}
\affiliation{\Jena}

\author{Bernd \surname{Br\"ugmann}}
\affiliation{\Jena}

\author{Frans \surname{Pretorius}}
\affiliation{\Princeton}

\date{September 12, 2012}

\begin{abstract}
Neutron star binaries offer a rich phenomenology in terms of
gravitational waves and merger remnants. However, 
most general relativistic studies have been performed for nearly
circular binaries, 
with the exception of head-on collisions.
We present the first numerical relativity investigation of mergers of
eccentric equal-mass neutron-star binaries that probes the regime
between head-on and circular.
In addition to gravitational waves generated by the orbital motion, we
find that the signal also contains a strong component due to stellar
oscillations ($f$-modes) induced by tidal forces, extending a
classical result for Newtonian binaries. The merger can lead to rather
massive disks on the order of 10\% of the total initial mass. 
\end{abstract}


\maketitle


\section{Introduction.} 
Binary neutron star (NSNS) mergers are among the most promising
sources of gravitational waves (GWs) for ground-based interferometers,
as well as plausible candidates for the central engine of short-gamma-ray-burst (sGRB).
One of the main tools for studying NSNS mergers is numerical
relativity, which experienced a dramatic development in the last ten
years achieving important results in
NSNS and recently also for black hole-neutron star (BHNS)
simulations, e.g.~\cite{Due09,lrr-2011-6,BauSha10x}.

In numerical relativity NSNS mergers have been almost exclusively
studied for circularized initial data. The exceptions are head-on
collisions~\cite{JinSue06,KelRezRad10},
and the low eccentricity ($e\approx0.2$) orbits of~\cite{AndHirLeh08}
also recall the residual eccentricity of quasi-circular data~\cite{Mil03}. 
(See~\cite{ShiNakOoh93} for a nearly head-on Newtonian
simulation with radiation reaction.)
So far, there are no numerical studies exploring
the range from small to large eccentricity.
One reason for this is an expectation that the dominant population
of NSNS mergers in the universe results from evolution of primordial
stellar binaries. For such systems, residual eccentricity at
the time of merger is expected to be low. For example,
a recent investigation suggests that only between $0.2\%$ and $2\%$ of this
class of mergers detectable by Advanced LIGO/VIRGO will have eccentricity
$e>0.01$ (and $e\lesssim0.05$)~\cite{KowBulBel11}. 

However, recent studies have suggested there may be a significant
population of compact object binaries formed via dynamical capture in
dense stellar environments~\cite{oleary,LeeRuiVen09}, and a large
fraction of these would merge with high eccentricity.
In \cite{LeeRuiVen09} the number
of dynamical capture NSNS mergers in globular clusters (GCs) was calculated, finding a 
redshift dependent rate that peaks
at $50$/yr/Gpc$^3$ at $z=0.7$, decreasing to $30$/yr/Gpc$^3$ today.
This is large enough to account for a significant fraction
of sGRBs, assuming NSNS mergers are their progenitors. 
However, the calculated merger rate is quite sensitive to the fraction $f$ of NSs in the core, scaling
as $\sim f^2$, and their model (using M15 as the prototype GC) did not account
for NS loss due to natal kicks. A recent simulation
of M15 by~\cite{2011ApJ...732...67M}
fit to observations and assuming a modest NS
retention fraction of $5\%$, found roughly $1/4$ fewer NSs within the central $0.2$ pc
compared to an earlier study that did not account for kicks~\citep{2003ApJ...585..598D},
implying the rates of~\cite{LeeRuiVen09} could be overestimates by an order of magnitude due
to this effect.
On the other hand, the $5\%$ retention fraction used in~\cite{2011ApJ...732...67M}
may be too low, as observations of GCs suggest it could be as 
large as $20\%$~\citep{2002ApJ...573..283P}. Furthermore, the model of~\cite{LeeRuiVen09} did not
take into account other channels that could lead to high eccentricity binary merger within a Hubble
time, such as Kozai resonance in a triple system~\citep{Thompson:2010dp}.
Observations of
sGRBs have also suggested there are different progenitors for sGRBs
that exhibit so called extended emission and those that do
not~\cite{Norris:2011tt}; an obvious speculation of the cause of the
bimodality is primordial vs.\ dynamical capture NSNS mergers. 

Although the issue of event rates is unsettled, we expect
smaller but not negligible rates for eccentric binaries
in the overall population. 
We therefore posit their existence, and ask what theory predicts for
the evolution and the gravitational waves of eccentric NSNS binaries.

Eccentric mergers can lead to a rich phenomenology in GWs and the
merger remnant.
The properties of the accretion disk formed as a product of the
merger are so far poorly understood, even though there is the
possibility of massive disk production, $M_{\rm d}\sim10\%$ of the
total initial binary mass. Recent results~\cite{SteEasPre11} for BHNS
systems indicate a strong variability in properties of the merger remnant 
as a function of the eccentricity.
The gravitational waveforms significantly differ from the chirp signal
of quasi-circular inspirals that feature slowly increasing amplitude.
During eccentric orbits, each periastron passage leads to a burst of
radiation, first studied using Newtonian orbits together with leading
order relativistic expressions for radiation and the evolution of orbital
parameters~\cite{PetMat63,Tur77a}, and more recently with numerical
simulations of BHBH and BHNS systems in full general
relativity~\cite{PreKhu07,SpeBerCar07,GolBru09,SteEasPre11,GolBru12}.
An important Newtonian result concerning NSNS (or BHNS) is that
eccentricity leads to tidal interactions that can excite oscillations
of the stars, which in turn generate their own characteristic GW
signal~\cite{Tur77b} (see also \cite{KokSch95,LeeRuiVen09}).
In some cases the GWs can be dominated by these non-orbital
contributions~\cite{Tur77b}. Neither the orbital nor stellar GWs
have been studied so far for eccentric NSNS orbits in general
relativity.

In this work we present the first numerical relativity investigation of
(highly) eccentric NSNS mergers (see
Refs.~\cite{LeeRuiVen09,Rosswog:2012wb} for studies of similar systems
in Newtonian gravity).  We consider an equal-mass binary at fixed
initial separation and vary the initial eccentricity.  Three models are
discussed that are representative of the different orbital phenomenology
observed: direct plunge, merger after a close encounter, and multiple
encounters.  We analyze the emitted GW and the dynamics in the matter,
and characterize the basic properties of the merger remnant.


\section{Numerical Method.}
We performed numerical simulations in 3+1 numerical relativity
solving the Einstein equations coupled to a perfect fluid matter model
(no magnetic fields). We employed the
\bam~code~\cite{BruTicJan03,BruGonHan06}, which we recently extended
to general relativistic hydrodynamics~\cite{ThiBerBru11}.
Referring to~\cite{BruGonHan06,ThiBerBru11} for details and further references,
we solve the BSSN formulation in the moving puncture gauge coupled to
matter in flux-conservative form.
Metric and matter fields are discretized in space on 3D
Cartesian meshes refined with the technique of moving boxes. 
Time integration is performed with the method-of-lines using a 3rd
order Runge-Kutta scheme.  Derivatives of metric fields are
approximated by fourth-order finite differences, while a
high-resolution-shock-capturing scheme based on the
local-Lax-Friedrich central scheme and the
convex-essentially-non-oscillatory reconstruction is
adopted for the matter.  Neutron star matter is modeled with a
polytropic equation of state with adiabatic index
$\Gamma=2$. GWs are extracted at finite coordinate
radius, $r\sim 90M$ ($180M$ for Model~3), by using the Newman-Penrose curvature scalar,
$\Psi^4$, where $M=2.8\Msun$ is the gravitational mass of the system.

Although the problem of posing initial data to the Einstein equations
by solving the constraints is in principle well understood, there 
is no ready-made prescription for eccentric NSNS that is of 
interest here (although see the recent work in Refs.~\cite{East:2012zn,Mol12}). 
We postpone the solution of the constraint equations
and instead work with the following approximation.
Initial data are set by superposing two boosted non-rotating stars
with the same gravitational mass $1.4\Msun$ at apoastron.
Specifically, we 
(i)~compute a relativistic spherical star configuration (TOV) of mass~$M_\star=1.4\Msun$;
(ii)~obtain two boosted star configurations by performing the Lorentz
transformation along the $y$-axis with parameters $\xi=(0,\pm\xi_y,0)$;
(iii)~place the star centers at coordinate locations $(x_\pm,y,z)=(\pm25~M,0,0)$, respectively;
(iv)~add the 4-metric fields of the two spacetimes and subtract the flat metric in
order to enforce asymptotical flatness;
and (v) derive 3+1 decomposed initial data.
The initial spatial metric is not conformally flat.
The matter initial data are
approximately irrotational and represent orbiting neutron stars near apoastron.
The maximum constraint violation (at the
center of the star) decays as $1/d$ as expected for the relatively
large initial proper separation $d\sim54.7M$ (compared to typical 
separations of quasi-circular initial data, e.g.~Refs.~
\cite{ThiBerBru11,BerNagThi12}). 
The constraint violation is at the level of the truncation error of the 
evolution scheme (see also the discussion in Refs.~\cite{AndHirLeh08,SteEasPre11}).

We performed a series of simulations for different initial boosts
(i.e.~Newtonian apoastron velocities), $\xi_y\in[0.01,0.05]$. 
The eccentricity arising from these choices cannot be unambiguously 
quantified in GR. The Newtonian equivalent of these 
configurations would lead to eccentric elliptic orbits with eccentricities 
as listed in Tab.~\ref{tab:remnant}. Newtonian circular orbits are obtained for
$\xi_y\simeq0.07$. The evolution 
of the orbital separation and the GW frequency suggest that the Newtonian 
values overestimate the eccentricities.

The grid consisted of five refinement levels with maximum spatial
resolution $h\!\sim \!0.15-0.2\Msun$ and a Courant-Friedrichs-Lewy
factor of $0.25$. We have checked that varying the box size, $B\!\sim
\!200-400M$, has only minor effects on the orbits and the waves. For
$\xi_y\lesssim0.020$, within $1000M$ a merger and formation of a black
hole occurred. Selected models were continued for $\sim \!\! 700M$
($\sim \! 10ms$) after the merger in order to follow the accretion
process on to the final black hole.  During evolutions the largest
violation of the rest-mass conservation occurs before collapse, $\Delta
M_0/M_0\sim0.01$. Similarly the ADM mass was conserved up to $1\%$. The
consistency of the results was assessed by convergence tests.

\section{Orbital dynamics.}
\begin{figure}[t]
  \centering
  \includegraphics[width=0.49\textwidth]{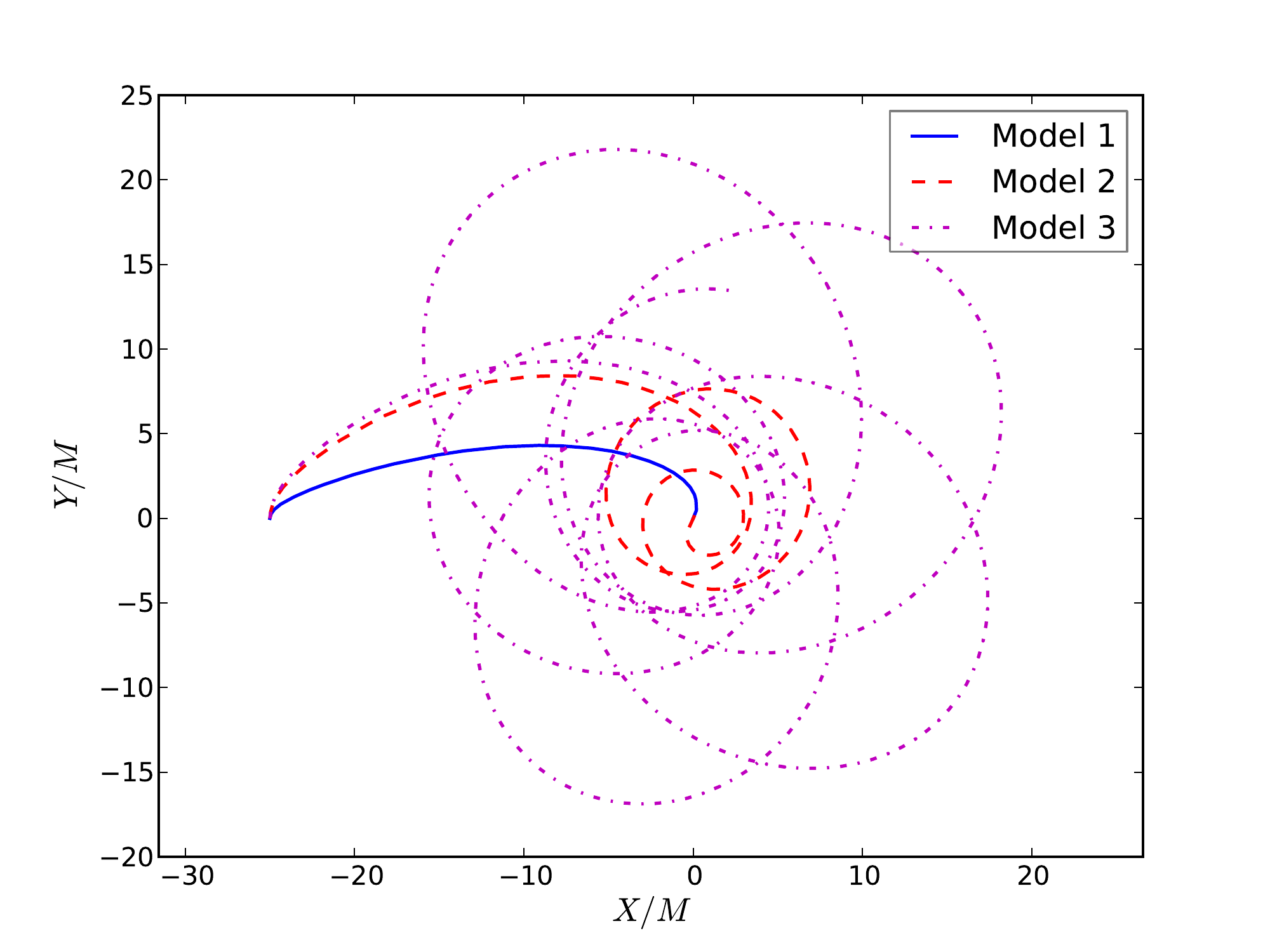}
  \caption{ \label{fig:startracks} 
    Star tracks for Model~1,~2, and~3. 
    The tracks of the second star can be inferred from symmetry.}
\end{figure}

We begin by reporting the orbital dynamics in selected models. 
See Fig.~\ref{fig:startracks} for the star tracks as computed 
from the minimum of the lapse function~\cite{ThiBerBru11}. 

\begin{figure}[t]
  \centering
   \includegraphics[trim=10mm 160mm 20mm 30mm, clip, width=0.49\textwidth]{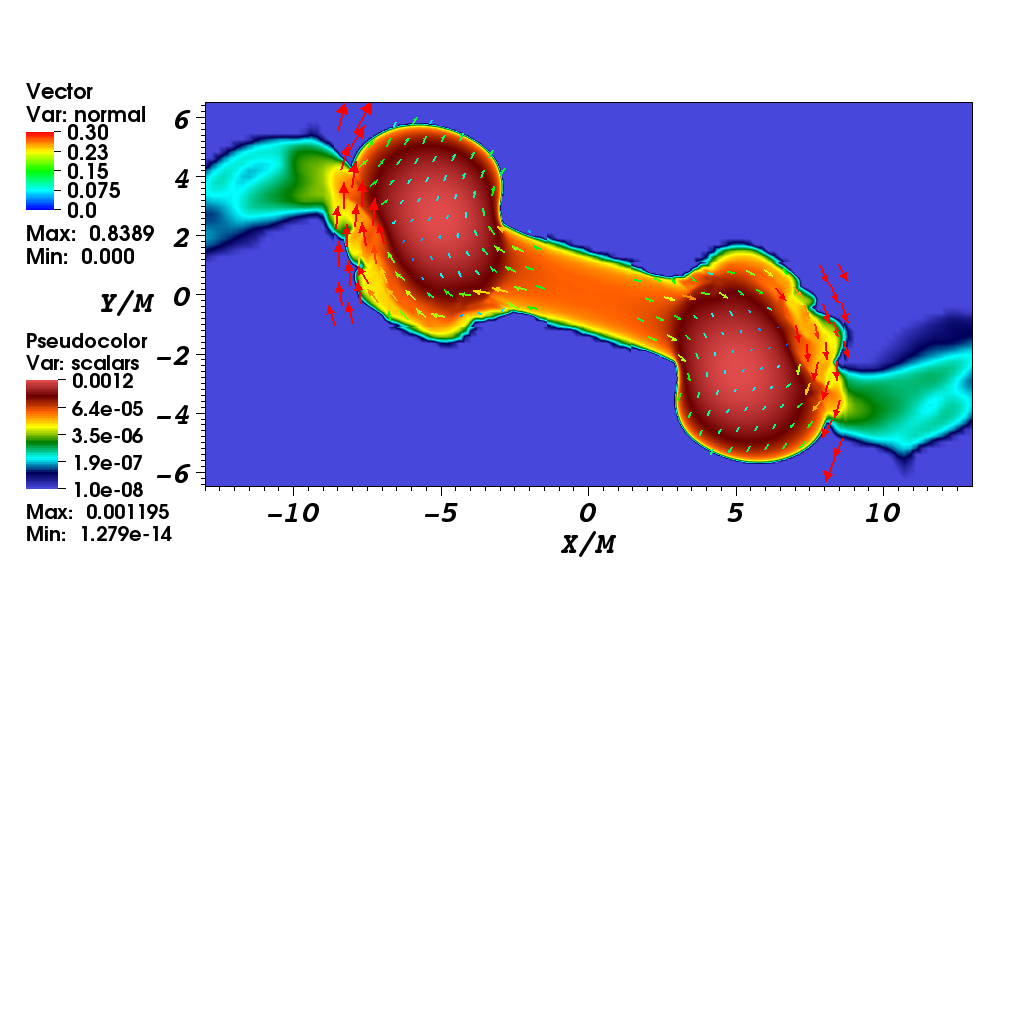}
  \caption{ \label{fig:rhov_xy_1enc}
    Model~2 at $t=553M$, shortly after the stars have touched and separated again.
    The rest-mass density (log-scale) and the three-velocity in the orbital plane are shown.}
\end{figure}

In Model~1 ($\xi_y=0.010$) the stars move almost directly towards 
each other, collide, then undergo prompt collapse to a black hole at
$t=423M$ with dimensionless spin parameter  
$a_{\rm BH}\sim0.58$ and mass $M_{\rm BH}\sim2.8\Msun$.
The resulting disk has negligible rest-mass, $M^d \lesssim 10^{-7}M_0$, 
which is at the level of the artificial atmosphere used for the
numerical treatment of vacuum for the matter. 

In Model~2 ($\xi_y=0.020$) the stars come into contact but survive the first 
encounter. During contact matter is exchanged between the outer layers of 
the stars, see Fig.~\ref{fig:rhov_xy_1enc}; 
The density weighted Newtonian vorticity is 10 
times larger than for an irrotational NSNS binary. 
During a second encounter they merge forming a black hole at $t=975M$
with initial $a_{\rm BH}\sim0.79$ and mass $M_{\rm BH}\sim 2.2\Msun$. In contrast
to Model 1, there is a massive disk with $M^d=(0.15 \pm 0.02)\Msun$
(the error estimate is based on five resolutions between 
$h\sim0.15-0.2\Msun$), 
measured $100M$ after the formation of an apparent
horizon. We find larger masses up to $M^d \sim 0.27\Msun$ for intermediate 
models which are not discussed here in detail, see Tab.~\ref{tab:remnant}. 
The accretion has increased 
the mass and spin to $M_{\rm BH}\sim2.64\Msun$ and $a_{\rm BH}\sim 0.81$. 
Prior to the onset of merger, for a time of $\approx 100M$ the binary approaches
a nearly circular orbit where the separation remains close to a constant $\sim 5M$.
This 
suggests a transition through a whirl regime as in
BBH systems~\cite{HeaLevSho09,PreKhu07,SpeBerCar07,GolBru09,SteEasPre11,GolBru12}. 
After the merger we observe quasi-periodic oscillations in
the accretion flow at a frequency $\nu_{QPO}\sim 0.5kHz$ 
in the vicinity of the black hole.

In Model~3 ($\xi_y=0.022$) multiple close passages with no matter
exchange are observed.  The two bodies do not merge during a
simulation time of $t=6400M$.  The orbits show a significant 
periastron precession, leading to an overall rotation of the
quasi-elliptical orbit of almost $90$ degrees per encounter.
Although the precession is highly relativistic,
these encounters occur well outside the whirl regime.

\begin{figure}[t]
\centering
\includegraphics[width=0.49\textwidth]{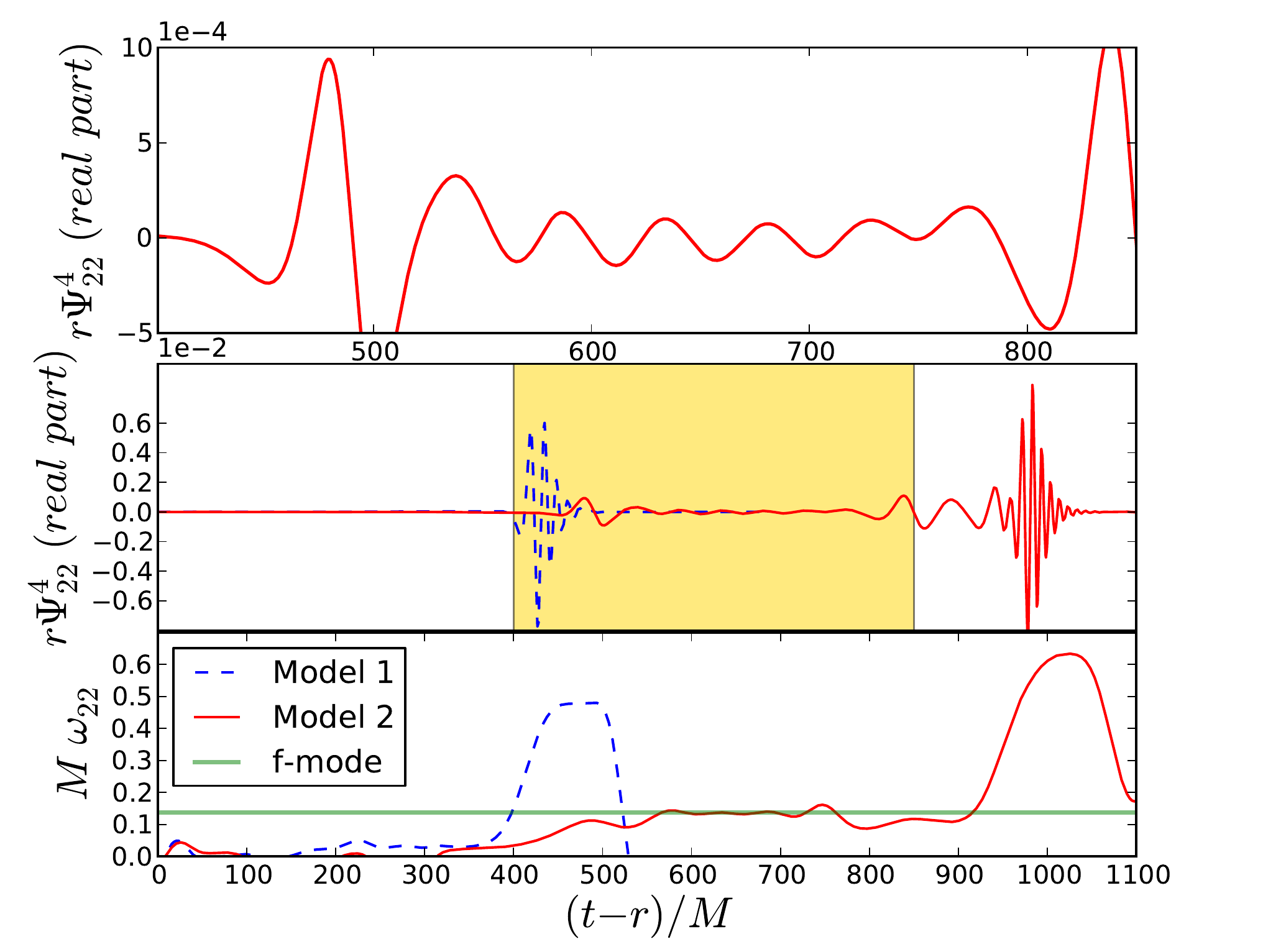}
\caption{ \label{fig:waves} 
  Real part of the $r\,\Psi^4_{22}$ waveforms (central panel) and instantaneous 
  gravitational wave frequency (bottom panel), $M\omega_{22}$, as a function of 
  retarded time, $(t-r)/M$, for Models~1 and~2. The top panel shows a  
  zoom in on the shaded region in the central panel for Model~2. 
  The horizontal line in the bottom panel marks the perturbative $f$-mode frequency.
}
\end{figure}

\begin{figure}[t]
  \centering
  \includegraphics[width=0.49\textwidth]{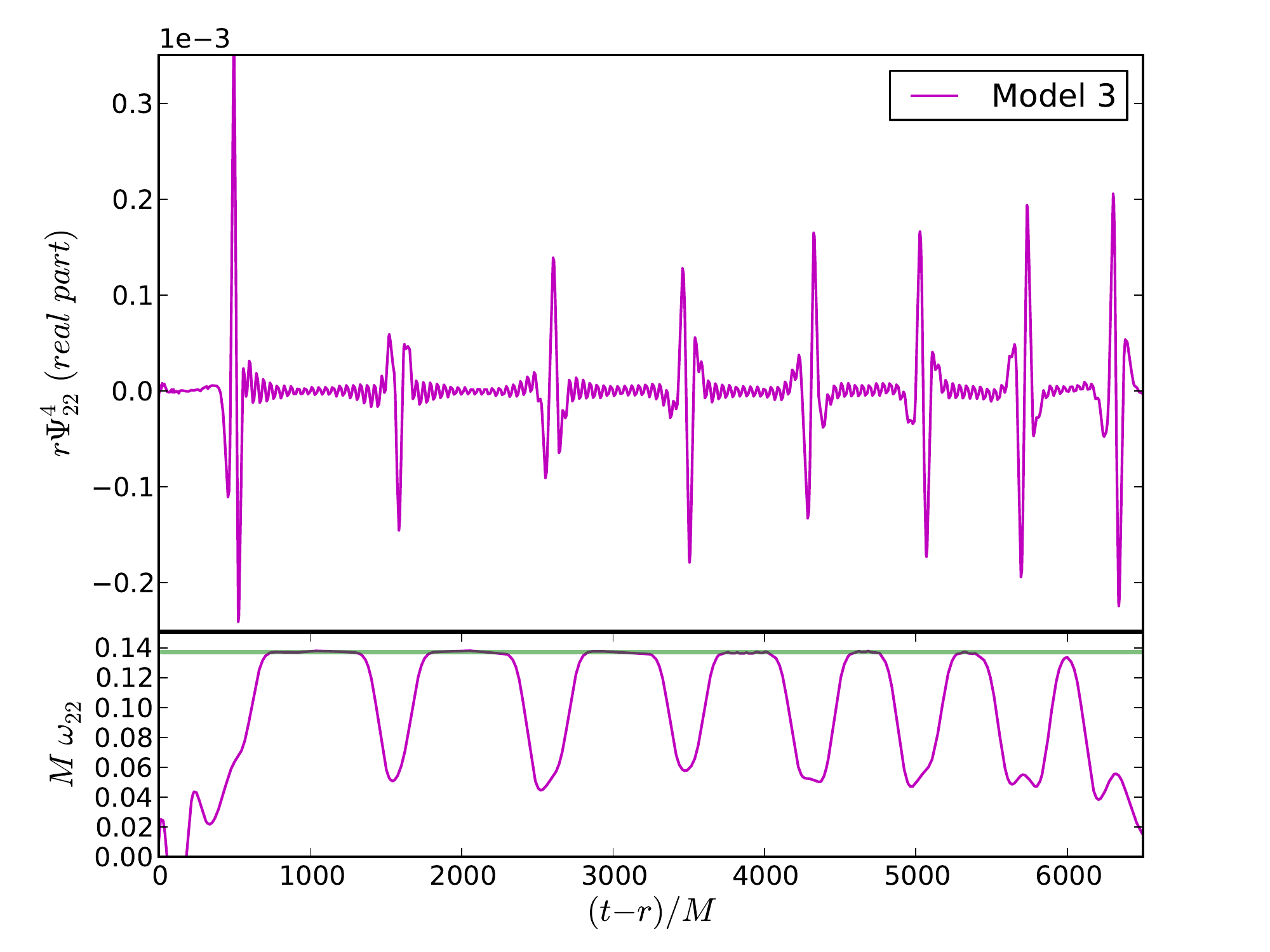}
  \caption{ \label{fig:omega} 
    Same as the central and bottom panels of Fig.~\ref{fig:waves} but for Model~3.}
\end{figure}


\section{Waveforms.}
The orbital motion of each model results in a characteristic
gravitational wave signal. 
Figs.~\ref{fig:waves} and \ref{fig:omega} display the $\ell=m=2$ mode, $\Psi^4_{22}$, 
as well as the corresponding instantaneous gravitational wave frequency, $M\omega_{22}$,  
of the curvature scalar, which represents the main emission channel for the 
binary signal. 

Waveforms from Model~1 are characterized by a peak at the
merger and the subsequent quasi-normal-mode (QNM) ringing of the final
black hole. The GW frequency of the QNM is $M\omega_{22}\sim0.48$,
compatible with the fundamental QNM frequency of the Kerr black hole 
computed from the apparent horizon parameters, e.g.~\cite{Pre07}. 
In fact, we find the value $\nu_{\rm QNM}\sim5.9kHz$, which differs
by $6\%$ from the value obtained from the GW.
Waveforms from Model~2 show a burst at retarded time $t-r\sim500M$ 
related to the orbital dynamics~\cite{Tur77a}, followed by a feature 
between $t-r\sim [800M-900M]$ akin to a transition through a whirl 
phase and the final merger signal around $t-r\sim1000M$. 
In this case the QNM frequency is higher, $M\omega_{22}\sim0.632$
($\nu_{\rm QNM}\sim7.6kHz$, $2\%$ discrepancy).
The waveforms from Model~3 exhibit several bursts corresponding to the
periastron passages.

The key feature is that between the bursts one can observe high
frequency signals in several GW-modes (e.g.~for Model 2 at
$t-r\sim[550M-750M]$). These are absent in the black hole case, but
similar to Newtonian results \cite{LeeRuiVen09}, and qualitatively
similar to BHNS mergers~\cite{LofRezAns06,SteEasPre11}.
This signature is progressively suppressed 
at lower eccentricities (larger $\xi_y$)
and not observed for $\xi_y=0.05$.  
Referring to Fig.~\ref{fig:omega}, the $(2,2)$ GW frequency is
dominated by several plateaus compatible with the $f$-mode frequency
of the non-rotating star in
isolation~\cite{DimSteFon05,BaiBerCor08,GaeKok10}, 
$M\omega_{\rm f}\sim0.137$ ($\nu_{\rm f}\sim1.586kHz$), 
interrupted briefly by the close encounter bursts.  
In Model~2, the signal is mainly in the $(2,2)$ multipole; the $(2,0)$ multipole (not shown) also contains the signature (weaker in amplitude by a 
factor five). In contrast, in Model~3 the amplitude of the $(2,2)$ 
multipole is smaller in amplitude by a factor of three than the $(2,0)$ 
multipole, which is the main emission channel between the bursts 
(encounters).

The natural interpretation is that these GWs are produced by
oscillations of the individual stars, which are tidally induced by the
companion. Physically, the expectation is that the approximately
head-on, zoom-part of the orbit is responsible for exciting
axisymmetric $m=0$ modes, while in particular $m=2$ modes should arise from
the approximately circular, partial whirl near periastron.
The  relative amplitude of the modes indicates the effectiveness of 
the different phases of the binary interaction to excite certain modes.
Indeed the Newtonian analysis of Turner~\cite{Tur77b} indicates that a
sufficiently close periastron passage (i.e.\ $e\lesssim 1$ for fixed
apoastron) exerts a pulse-like tidal perturbation which excites the
axisymmetric $f$-modes of the star, leading to a GW waveform dominated
by the star oscillations rather then by the emission due to the
orbital motion. 
Note that the phenomenon is different from a resonant tidal
excitation, 
see e.g.~\cite{KokSch95,HoLai98}, which instead refer to the circular
motion case.

\begin{figure}[t]
  \centering
  \includegraphics[trim=0mm 70mm 0mm 60mm, clip, width=0.49\textwidth]{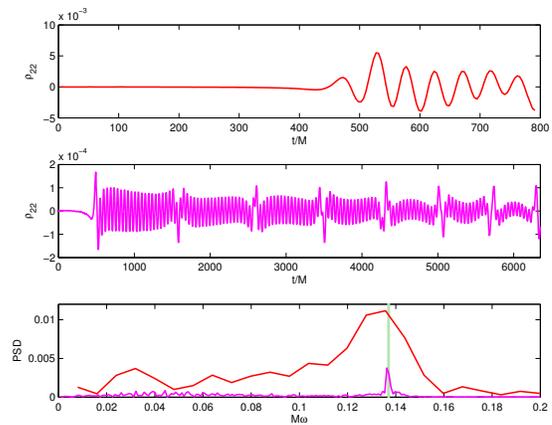}
 \caption{
   Mode analysis. 
   The panels display the $\ell=m=2$ spherical-harmonic projection of the rest-mass
   density  of one of the stars (Model~2 top, Model~3 center), 
   and its power spectral density (PSD, bottom), where $\nu_f$ is marked.
  {The green shaded area spans the perturbative values of the mode frequencies of star configurations A0-A1 in \cite{DimSteFon05} (right-left).}
}
  \label{fig:rholm} 
\end{figure}

If these waves indeed correspond to stellar $f$-modes, they should be
detectable as oscillations of the stellar matter.
Consider the projections of the
rest-mass density of one of the stars onto the spherical harmonics,
e.g.~\cite{BaiBerCor08},  
$
\rho_{\ell m}\equiv\int\,Y^*_{\ell m}\rho\,d^3x,
$
and perform an analysis of the spectrum.
Results are reported in Fig.~\ref{fig:rholm} for Model~2 and~3, where
the $\rho_{22}(t)$ projections 
and their power spectral density (PSD) are shown.
In both these (and in other intermediate) models we can
clearly identify frequencies compatible with the linear $f$-mode,
together with secondary peaks presumably due to nonlinear couplings
and/or non-axisymmetric modes.
Similar results are obtained for the projection $\rho_{20}$, while in
this case the signal is strongly modulated by the orbital motion and
contains a signature of the radial ($\ell=m=0$) $F$-mode of the star
from the beginning of the simulation.  

In \cite{DimSteFon05} the $f$-mode frequencies for different stellar 
models are given. In particular, the $f$-mode frequency is sensitive to 
the rotation of the star. Note that, given an uncertainty in the frequencies 
obtained from our simulations, we can derive an upper limit on the 
rotation of the star.
In Model 2 the peak in the PSD corresponding to the $f$-mode frequency is 
too broad for a restrictive upper limit. The broadening is caused by the 
limited number of cycles due to the short signal.
In Model 3 there are more cycles leading to a more accurate determination of the frequency. 
The latter lies between model A0 (non-rotating case) and A1 in \cite{DimSteFon05}. 
We conclude that the individual neutron stars have lower values of rotational/potential 
energy $T/W$ than model A1, which is given as $T/W=0.02$.
$r$-modes are thus not expected, but they may be present in the outer
layers of Model~2; $g$-modes are absent here by construction.

\begin{figure}[t]
  \centering
  \includegraphics[width=0.49\textwidth]{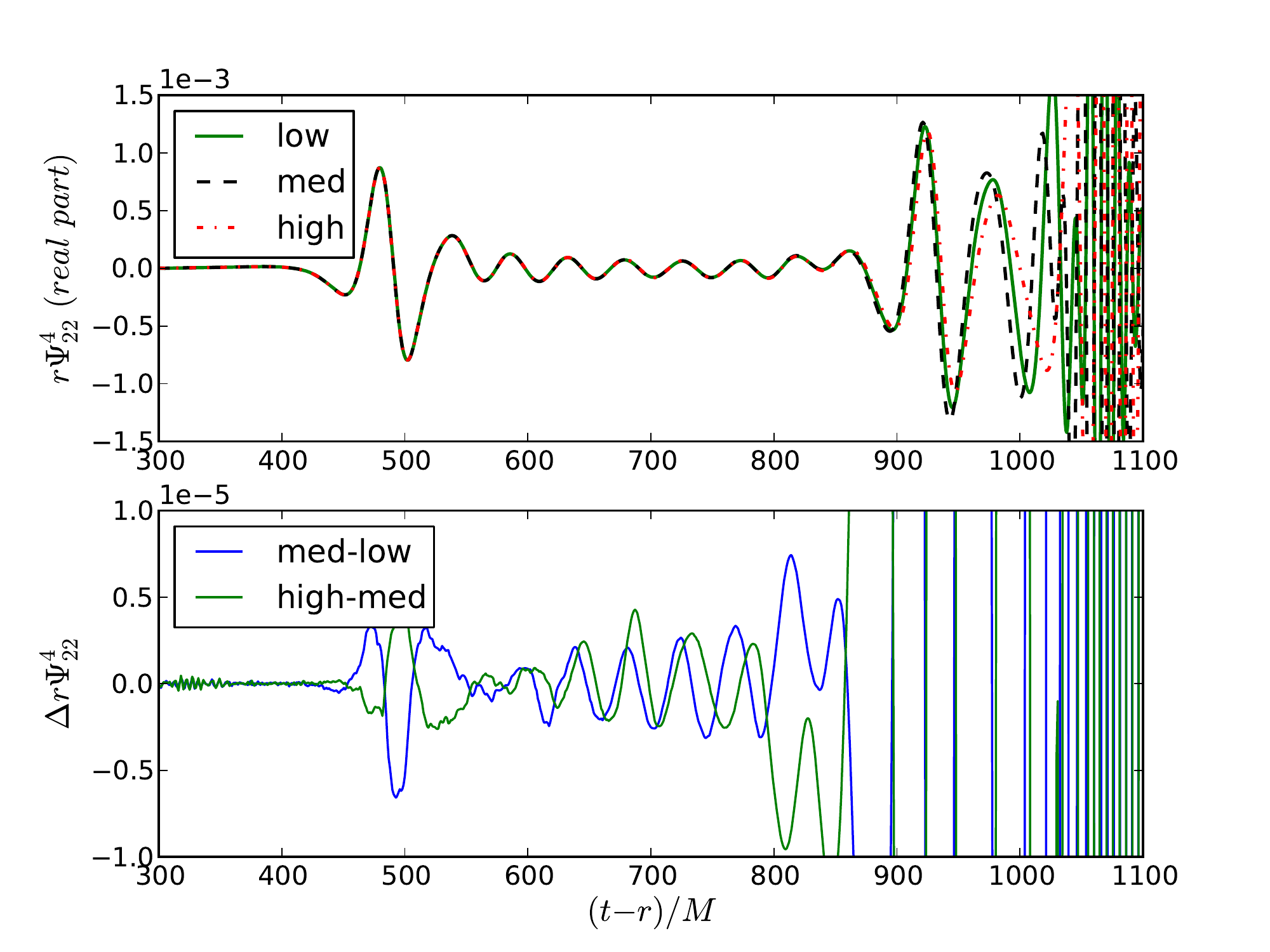}
  \caption{ \label{fig:gw_res} 
    Real part of the $r\Psi^4_{22}$ waveforms of Model 2 at three different resolutions. 
    The differences are two orders of magnitude lower than the $\nu_f$ signal. At $t \gtrsim 900$ 
    the phase error becomes large due to the highly sensitive whirl motion.
  }
\end{figure}

The main observation is that $f$-modes are excited during the encounter, i.e.\
the $f$-mode is present in all models that survive the 
first encounter. The value for $\nu_f$ remains unchanged for 
different resolutions (see Fig.~\ref{fig:gw_res}) and different initial data, 
demonstrating the robustness of our finding.

\begin{figure}[t]
  \centering
  \includegraphics[width=0.49\textwidth]{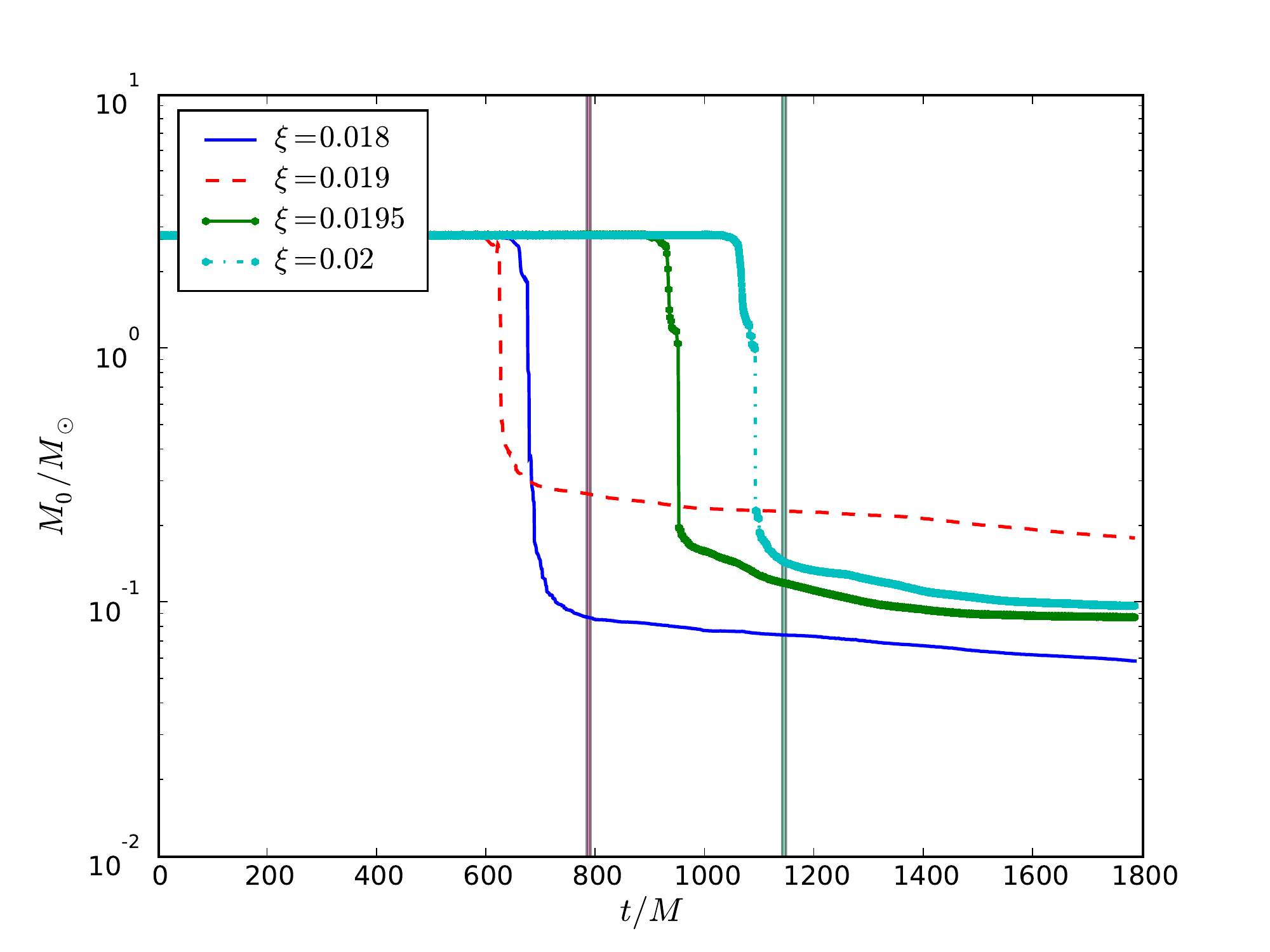}
  \caption{ \label{fig:restmass} 
    Rest-mass density evolution of Model 2 and selected intermediate models. The vertical lines denote $t_d$ for $\xi = 0.019$ and $\xi = 0.02$.
}
\end{figure}

Finally, we discuss the properties of the merger remnant and the
  surrounding accretion disk for models with $\xi_y$ between those of
  the previously discussed Models 1-3.  In Tab.~\ref{tab:remnant}
  initial parameters of the different models are listed together with
  merger time $t_{AH}$, horizon mass $M_{BH}$, and spin
  $a_{BH}$. The superscripts $0$ and $d$ denote times $t_{AH}$ and
  $t^d\equiv t_{AH}+100M$, respectively. The latter time is
  empirically motivated by an initial phase of rapid change in the
  merger properties. We find that after $t^d$ the remnant and the disk
  are changing much more slowly. See Fig.~\ref{fig:restmass} for the evolution of the rest-mass as a function of time for selected models. 
  The results are in qualitative agreement with corresponding results in which constraint-satisfying initial data are used \cite{EasPre12}. 
  We find final spins of $0.75 < a^d_{BH} < 0.81$ for all models (except Model 1) and disk masses
  $M^d \gtrsim 0.18M_\odot$ for $\xi_y \geq 0.019$. 
  As found in \cite{EasPre12}, grazing encounters exhibit larger disk masses. 
  In the early
  post-merger stages that we simulate, the accretion rate can be fit to
  an exponential function, which allows us to quote a characteristic
  accretion time scale $\tau$. These values are given in the last
  column of Tab.~\ref{tab:remnant} and range from $4ms$ to $30ms$.

\begin{table}[t]
  \caption{ \label{tab:remnant} 
    Summary of the merger remnant properties.
    Columns: initial data boost parameter, 
    Newtonian eccentricity of the initial data, 
    mass and spin of the final black hole at two different times as estimated from the
    apparent horizon, disk rest-mass, and accretion time. Errors are about $1~\%$.}
   
 \begin{center}
    \begin{tabular}{cc|cccccccccc}
      \hline 
      $\xi_y$& $e$   &$t_{\rm AH}/M$&$M_{\rm BH}^0$& $M^d_{\rm BH}$ & $a^0_{\rm BH}$ & $a^d_{\rm BH}$ &
      $M^{\rm d}/M_\odot$ & $\tau$[$\times10^3$]\\
      \hline  								
      0.01   & 0.978 &  423 & 2.82 & 2.81& 0.58 & 0.57 &$<\!10^{-7}$& -\\
      0.017  & 0.937 &  669 & 2.05 & 2.78& 0.64 & 0.76 & 0.004& 0.56M \\ 
      0.0175 & 0.933 &  658 & 2.06 & 2.75& 0.66 & 0.75 & 0.05 & 1.84M \\ 
      0.0185 & 0.925 &  692 & 2.13 & 2.73& 0.74 & 0.78 & 0.07 & 0.93M \\ 
      0.019  & 0.921 &  616 & 2.14 & 2.54& 0.71 & 0.75 & 0.27 & 2.72M \\ 
      0.0195 & 0.917 &  643 & 2.02 & 2.60& 0.72 & 0.79 & 0.20 & 0.47M \\ 
      0.02   & 0.913 & 1077 & 2.05 & 2.61& 0.70 & 0.81 & 0.18 & 1.81M \\ 
      \hline
    \end{tabular}
  \end{center}
\end{table}


\section{Conclusion.}
Mergers of eccentric NSNS binaries, 
although expected to be rare,
could be very interesting sources for third generation GW detectors,
in particular given the (in some respect) surprisingly strong and
clear signal from orbit-induced stellar oscillations (OISOs).
We find disk masses on the order $M^d \sim 0.2$ and remnant 
BH spins $a_{BH} \sim 0.8$.
Due to the sensitive dependence of the binary evolution on 
orbital parameters, signals from OISOs could pose constraints on the 
equation of state.
This motivates an extension of this work in the future to a more realistic 
equation of state 
including a quantification of the shock heating and the 
possibility of the formation of a hypermassive neutron star.
Our study 
raises the question of to what extent a NS crust could tolerate such 
strong deformations. As also pointed out in \cite{EasPre12} 
the crust may fail during the inspiral.

This study is preliminary in many aspects, but it represents a first step 
toward building GR models of eccentric binary neutron stars. 
We consider improved initial data and a more realistic matter model 
(see \cite{EasPre12,Mol12}) to be the most important extensions.
The present results are likely to be highly dependent on the
compactness of the stars and on the mass ratio, thus these cases
should be studied as well. Furthermore, even larger disk masses are
expected for the unequal mass case which could be interesting for sGRB
astrophysics. More work is required in these directions together with
a detailed understanding of the eccentric NSNS population in the
universe.


\section{Acknowledgments.}  
We gratefully acknowledge helpful discussions with William East,
Branson Stephens, and Kostas Kokkotas.  
This work was supported in part by DFG SFB/Transregio~7.
RG was supported by DFG GRK 1523,
in particular during a stay at Princeton.
FP was supported by NSF Grant No.~
PHY-0745779 and the Alfred P. Sloan Foundation.
Computations were performed on JUROPA (J\"ulich) and at the LRZ
(Munich).



\bibliography{refs,tovtov,refsextra}


\end{document}